%Paper: hep-th/9205061
%From: AEQJ000@ucsvax.cs.umass.edu
%Date: Mon, 18 May 92 14:46 EST

\input phyzzx
\date{May, 1992}
\Pubnum={UMHEP-371/revised \cr hep-th/9205061 }
\titlepage
\title{Quantum-Mechanical Scattering of Charged Black Holes}
\author{Jennie Traschen}
\address{Department of Physics\break
University of Massachusetts\break
Amherst, Massachusetts}
\andauthor{Robert Ferrell}
\address{Department of Physics\break
University of California\break
Santa Barbara, California}
\vfil
\abstract
We describe the quantum mechanical scattering
of slowly moving maximally charged black holes.  Our technique is to develop
a canonical quantization procedure on the parameter space
of possible static classical
solutions.  With this, we compute the capture cross sections for the
scattering of two black holes.  Finally, we discuss how quantization
on this parameter space relates
to quantization of the degrees of freedom of the gravitational field.
\endpage

\chapter{Introduction}

 	Here we study quantum mechanical interactions of
charged black holes.  In classical
general relativity there exist exact static
solutions for N maximally charged black holes; the black
holes can be placed anywhere, and
will remain at rest.  This suggests that for slowly moving
maximally charged black holes, the
spatial geometry at any time will be well approximated by the
static solution for the black
hole configuration at that time.  The classical solutions were worked
out [2] following this type
of adiabatic approach.  Now, the parameter or moduli space of possible
static solutions is a 3N dimensional
manifold, consisting of the positions of the N black holes.
Therefore, for the slowly moving
black holes, a path is traced out in moduli space as the 3-
geometry evolves to the 4-dimensional spacetime.

	In studying the classical
solutions it was found that the motion of the black holes was
governed by an effective
Hamiltonian for N point particles.  The approach here is to use this
Hamiltonian to evolve a Schroedinger
wave function, for the case of two black holes.  The wave
function is a function of the
positions of the two black holes, that is, the configuration space of the
wavefunction is the parameter
space of possible classical static solutions.  Hence the degree of
freedom being quantized is the proper distance between the two black holes.

	One can imagine that if
the full theory for quantum gravity were known, one could compute
the motion of maximally charged
black holes and then consider the slow motion, low energy limit.
Here, we can hope that we
are computing an approximation to the actual motion.  Of course, since
we do not know the theory of
quantum gravity we have no way of knowing how good an
approximation this is.  On
the other hand, we do know of some ways this approximation may fail.
In particular, the classical metrics
were assumed to have a particular form, corresponding to slow
motion.  However, in the full quantum
theoretic description of the system, there will surely be
excitation of degrees of freedom
which are not included in this model, i.e., the configuration space
for the true wave function must
include other degrees of freedom than those in the moduli space
considered here.  In our approximation we
will never see these.  This is similar to mini-superspace
models, and also quantized
non-linear sigma models (when viewed as models of approximately
constrained systems), which
suffer the same shortcomingQthere is no way excite modes which
are not in the approximation.[1]

	Perhaps the most interesting
issue is the meaning of the wavefunction for the geometry of
the spacetime.  In the
classical case the metric can be found by solving constraint equations, with
the black hole positions and
momenta as sources, after the motion of the black holes is known.
When the sources are
quantized, one can no longer speak of a position and velocity of the source,
and instead must speak in terms
of probabilities.  The question, then, is what does it mean for the
metric field equations to
have sources which are probabilities?  Our preferred interpretation, but by
no means the only one,
is to consider the probabilities as probabilities for finding different
spacetimes.  We discuss these issues in the final section.

\chapter{The System for Scattering of Maximally Charged Black Holes}

	We now review the work of Ferrell and Eardley.[2]
This section will conclude with the
classical Hamiltonian for
black holes.  In the next section we will discuss the quantization
procedure, e.g., the Hilbert
space of states, and the quantum mechanical operator which
corresponds to the classical Hamiltonian.

	The sources in our study
will be maximally charged black holes.  A maximally charged
black hole is an electrically charged
(Reisner-Nordstrom) black hole with charge $Q = \sqrt{G} /M$.  Such
a black hole is maximally charged, since
if $Q$ is increased but $M$ is held fixed, the event horizon
which is associated with
the black hole disappears and a naked singularity arises.

	The theory for charged
matter in curved space is assumed to be the coupled Einstein-
Maxwell field theory.  The action for this theory is [3]:
$$S={1\over {16\pi G}}\int \ ^{(4)} R -{1\over {8\pi}}\int
F_{ab}F^{ab} +\int A_a j^a +\Sigma _i \int m_i d\tau _i  \eqn\action $$
In this expression,$\ ^{(4)} R$ is the scalar curvature
of the spacetime, $ A_a$ is the electromagnetic four-
potential, $j^a$ is the
electromagnetic current and $F_{ab} = \partial _a A_b -\partial _b A_a$
is the electromagnetic field
strength.  The last term
is the action for a collection of free particles, and doesn't make sense for
black holes.  In deriving the classical
effective action, Eardley and Ferrell replace the singularities
with charged dust, with charge
density equal to the mass density.  So in performing the
manipulations the sources
are smooth.  At the end of the derivation, the limit is taken in which the
charge density goes to a
sum of point singularities.  This regularization scheme works because the
dust is maximally charged, and
hence in a static configuration.  Dust with charge less than the mass
would tend to collapse.  For
this paper we will not make the distinction between black holes and
the smoothed distribution used to represent them.
	G is Newton's gravitational
constant.  In this paper we will work in units where G = 1 and
c = 1.  In these units $\hbar = m_{Planck}^2$
where $m_{Planck}$ is the Planck mass.

	In non-relativistic theories
it is clear that there exists a static configuration of maximally
charged black holes,
since the condition for maximal charge is just the condition that the static
gravitational attraction exactly
cancel the electrostatic repulsion.  Remarkably, in the full Einstein-
Maxwell theory for gravitation
coupled to electricity and magnetism there also can be static
configurations of maximally charged black holes.

	The metric that
describes a system of N static maximally charged black holes (first
discovered by Majumdar[4] and Papapetrou[5]) is:
$$	ds^2 = - \Psi ^{-2} dt^2 + \Psi ^{2} \delta _{ij}
dx^i dx^j , \eqn\metric$$
where $\nabla ^2 \Psi = -4\pi \Sigma _a m_a \delta
(\bf{x} - \bf{x}_a)$
with the black holes at the points $x_a$ and the boundary condition
$\Psi = 1$ at infinity.  Here $\nabla ^2 $ is
the Laplacian on the $\bf{x}$ regarded as coordinates on flat $R^3$.  In this
coordinate system the hypersurfaces
t = constant intersect the singularities at the points $x = x_a$, and
the event horizons are
the same points (i.e., the event horizons are represented as points, but have
surface area $4\pi m_a$.)  The Maxwell 4-potential is
$$A = -(1 - 1/\Psi )dt .\eqn\gauge$$

	The space of possible Majumdar-Papapetrou
metrics is the 3N dimensional configuration
space of the positions of the
black holes--once the $x_a$ are known, $\Psi$ is known and therefore the
metric and scalar potential
are also known.  If the black holes are moving slowly then the solution
will trace out a trajectory
in the space of static solutions.[2,6]  These quasi-static solutions exist
because in the slow motion
approximation radiation can be neglected:[3]  The static forces cancel
exactly, and the only relevant
forces are the velocity dependent magnetic force and the gravitational
analog, the gravito-magnetic force.
As long as the velocities remain small, there will be very little
radiation.  In this approximation
then, the fields are fully fixed by the positions and velocities of
the matter sources.  (A
similar example of this approximation comes from electrodynamics in flat
space.  If the position of
a charge q at ${\bf r}_0$ is changing slowly, with ${\bf v}
= d{\bf r}_0/dt$, then the scalar
potential is approximately $\phi (r,t) \approx
q/ | r - r_0 (t) |$, and the vector potential is approximately
${\bf A} \approx q{\bf v}/|r - r_0 (t) |$.  The
space of field configurations is just the configuration space for the particle,
since a field configuration is
determined by the position and velocity of the particle.)

	To describe slowly moving
black holes, one looks for a metric and four potential of the
form
$$ds^2 =- {1\over{\Psi}}dt^2 +2N_i dx^i dt +\Psi ^2 \delta _{ij}
dx^i dx^j \eqn\metrictwo$$
$${\bf A} =-(1- {1\over {\Psi}}) dt +A_i dx^i \eqn\gaugetwo$$
with $\Psi$ as before, but now the
$x_a$ will be functions of time.  Both $N_i$ and $A_i$ will depend on the
velocities of the black holes.
However, since we are assuming velocities are small we can truncate
the field equations for $N_i$ and $A_i$ to first order in velocity.

	Using the Einstein constraint
equations, the fields $N_i$ and $A_i$ can be solved for in terms of
the source positions and velocities.
These expressions are substituted into the Einstein-Maxwell
action \action .  The resulting
effective Lagrangian depends only on the positions and velocities of the
sources.  The black hole limit
is shown to be well defined.  This is the Lagrangian for the
interaction of the black
holes.  For details see [2].  Finally, one finds that the Hamiltonian for two
black holes, with masses $m_1$ and $m_2$, is
$$H_{2-body} ={1\over {2M}}{\bf P\cdot P} + {1\over {2\mu}}g^{ab}
p_a p_b . \eqn\twobody $$
where ${\bf P}$ is the momentum of the
center of mass, $M = m_1 + m_2 ,\  {\bf r} = {\bf x_1 - x_2}$ is the relative
coordinate, ${\bf p}$ is the momentum
conjugate to that coordinate, and $\mu = m_1 m_2 /M$  is the reduced mass of
the system.  In the first term,
a flat metric is implied.  The metric in the second term is
$$g_{ab}= \gamma (r) \delta _{ab} \eqn\gab$$
$$\gamma (r) =1+{{3M}\over r} +{{3M^2 }\over {r^2 }}
+{{3\alpha M^3 }\over {r^3 }} \eqn\gamr$$
where $\alpha = 1 - 2\mu /M$.
(Recall that ${\bf R}$ is the position of the center of mass, and
$\bf{r}$ is the relative  position, hence these are
coordinates on the moduli space of solutions, not on spacetime.)

	The evolution of the
center of mass coordinate is just free particle motion.  The evolution
for the relative coordinate
can have one of two behaviors.  If the two black holes start at infinite
separation, then when they
interact they can either scatter back out to infinity, or they can evolve
toward zero separation, depending
on their angular momentum.  Although in the $r\rightarrow 0$ limit the
slow motion approximation
probably breaks down, it is reasonable to guess that in this case the
two black holes coalesce
into one black hole.[2,3]  We will call the two possible classes of orbits
scattering orbits or coalescence
orbits, respectively.  This completes the review of the classical
behavior.

\chapter{Solutions of the Schroedinger equation for
Charged Black Holes }

Now we will view the hamiltonian $H_{2-body}$, for the system of two
maximally charged slowly moving black holes, as defining the evolution
of a quantum mechanical system. Hence the system is described by a wave
function. As in the classical case, the center of mass degrees of freedom
$X_c$ separate, so that the total wave function can be taken to be of
the form $\Psi_{total} = \Psi ({\bf x} ,t) exp \ i (-E_c t/\hbar +
{\bf P}_c \cdot {\bf X}_c )$, where $E_c$ and ${\bf P}_c$ are the center
of mass energy and momentum, and ${\bf x}$ are the relative coordinates.
Suppose that at early times a wave packet is given which has support at
large relative coordinate $r$, so the two black holes are far apart.
As time evolves the black holes approach each other, that is the wave
packet evolves towards smaller $r$. Part of the packet will be scattered and
part will be absorbed; classically of course the black holes either
scatter or coalsce. Here we will compute the absorption coefficient as a
function of the angular momentum and energy of the wave.
The Schroedinger equation for the wave function of the relative coordinates
is
$$-{\hbar \over i} \dot \Psi ({\bf x },t) = H\Psi \ -(E_c -{{\hbar^2
P_c^2 }\over 2M} - M)\Psi \ \ , \eqn\schrod$$
where
$$H=-{{\hbar ^2}\over {2\mu }} g^{ab} \nabla _a \nabla _b \  +
\hbar ^2\xi{\cal R}  $$
and
$$g_{ab} =\gamma (r) \left( dr^2 \  +r^2 d\Omega ^2 \right)  \ , \
and \  \gamma (r)= 1+{3M \over r } \  +{{3M^2}\over {r^2}} \ +{{\alpha M^3 }
\over {r^3}} . $$
${\cal R}$ is the scalar curvature of the three-metric $g_{ab}$, and
$\alpha = 1-2\mu /M$. The Hilbert space of states can be taken to be
square integrable functions. We will see that the energy eigenfunctions
are square integrable as $r\rightarrow 0$. As $r\rightarrow \infty$,
the eigenfuntions become plane waves, so just as usual, one would
need to form wave packets in the free particle regime. When checking
that $H$ is hermitian, one finds that boundary term contributions
of probability flux cancel, between large $r$ and the horizon. This is
similiar to the cancellation for plane waves in a box.

We take the eigenfunctions to be of the form $\Psi _{qlm} = \psi _{ql}
(r) Y_{lm} (\Omega )$, so that $H\Psi =\ E\Psi $ implies taht the radial
wave functions satisfy
$$\psi ''(r) \  +{2\over r}\psi ' \  +\  {1\over 2}{{\gamma '}\over
{\gamma }}\psi '(r)
 - {{l(l+1)}\over {r^2}}\psi
\  -2\mu \xi R \gamma (r) \psi
=\  - q^2 \gamma (r) \psi (r) \eqn\wavefun$$
where
$$q^2 = {{2\mu E}\over {\hbar ^2}}.$$
It is useful to change to a new "tortoise''
coordinate $R$, which measures actual length
along the path. This casts the problem into the form
of a standard one-dimensional
quantum mechanical scattering problem, and the potential is better behaved.
Let $R=\int \sqrt{\gamma }dr$ and let $\chi =r\gamma ^{1/2} \psi $. Then
the Schroedinger equation \schrod\ becomes
$$\chi ,_{RR} +\  (q^2 -\   V)\chi =0 \eqn\radial $$
where
$$V= {1\over {2r\gamma ^3 }}\left( \gamma (r\gamma ,_r ),_r -\  r
\gamma '^2 \right) + \  {{l(l+1)}\over {r^2 \gamma }}
+\  2\mu\xi {\cal R} .\eqn\potone$$
For $\mu =0$ we have
$$V={3\over 2}{1\over {M^2}} {{(r/M)^2 }\over {(1+r/M)^5 }}
+{{l(l+1)}\over {M^2 }} {{r/M}\over {(1+r/M)^3 }}\eqn\pottwo   $$
Here $r$ is an implicit function of $R$.
The new coordinate $R$ ranges from $-\infty $ at the horizon
to $+\infty $; $R\simeq
r+\  {3\over 2} M ln {r\over M}$ for $r\rightarrow \infty $ and
$R\simeq -2\sqrt{\alpha} M (r/M)^{-1/2} $ as $r\rightarrow 0 $.
For the remainder of the paper we will treat the case $\xi=0$. We
find, valid for all $\mu$,
$$V\simeq {3\over 2}{M\over {R^3}} +\  {{l(l+1)}\over {R^2}} \ , \
for \  R\gg M \eqn\potential$$
$$V\simeq 24{{M^2}\over {R^4}}+\  {{4l(l+1)}\over {R^2}} \  ,\
for R\ll -M .$$
Hence the potential falls off rapidly both in the asymptotic free region
and as the horizon is approached. If we think of the problem as
the quantum mechanics of a particle on
a curved surface, the three-geometry on which the particle moves becomes,
as the horizon is approached,
flat space minus a three-dimensional wedge of solid angle $4\pi /4 = \pi $.
Hence the eifenfunctions behave like free particles on either side of the
potential barrier, $\chi \simeq e^{\pm iqR} $ as $R\to \pm \infty $.
We want solutions to \radial\ for the eigenfunctions with the boundary
conditions
that $\chi$ is purely a captured, left-moving wave as $R\rightarrow -\infty
$, so there is no flux out of the horizon. Then as $R\rightarrow \infty $
the solution will be the sum of an incident wave $e^{-iqR}$, normalized
to unit amplitude, and a scattered wave $Se^{iqR}$. Using the asymptotic
form of the potential \potential, we find
$$\chi _{ql} \simeq \ qR\lbrack (-i)^{l+1} h_l^{(2)} (qR) + \  S_{ql}
h_l^{(1)} (qR) \rbrack \  ,\  \  R\gg M \eqn\rbig$$
and
$$\chi _{ql} \simeq \  C_{ql} qR h_{\nu}^{(2)}(qR) \  ,\  \   R\ll -M
\eqn\rsmall$$
where $\nu =\sqrt{4l(l+1) +1/4} -1/2 $. Hence the fraction of captured
flux is $|C_{ql}|^2$ and the scattered flux is $|S_{ql}|^2 =1-|C_{ql}|^2 $.

\chapter{General Properties of the Solutions}

First let us consider the motion of wavepackets, so each eigenmode
evolves
like $$exp (-i \hbar q^2 t/ 2\mu ).$$
At early tines, let a wave packet
start in the asymptotic free region, $R\to\infty ,\  t\to -\infty $.
Then by looking at the point
of stationary phase, one sees that initially the center of the packet moves
inward according to $R \simeq -v_{\infty}t $, where $v_{\infty }=\hbar q/
\mu $ is the nonrelativistic (relative coordinate) velocity at infinity,
associated with the momentum $\hbar q $. At large times the wave packet has
split into two pieces. The center of one part continues towards $R\to -\infty$
as $R\simeq -2v_{\infty}t$; this is the captured flux,
corresponding to classical coalescence of the black holes.
Hence it takes an
infinite amount of $t$-coordinate time for coalescence. There is also a
scattered part of the packet, propagating as $R\simeq v_{\infty } t$.
Hence the
motion of the center of the packet is like the classical solution.
The motion along the classical path
depends only on a rescaled time, $s=v_{\infty } t
$, and so the parameter $v_{\infty }$ scales out of the problem.
Hence the classical solutions depend
only on the impact parameter $b$. However,
quantum mechanically the quantities of physical interest such as the
scattering coefficient, will depend on $v_{\infty }$ (or equivalently, $q$
) as well as the analogue of $b$. Indeed, from
the eigenfunction equation \radial\ we see that the eifenfunctions depend on
the two dimensionless parameters $qM$ and ${b\over M}$, where
$${b\over M} ={{\sqrt{l(l+1)}}\over {qM}}
 \eqn\impact$$
or equivalently on $l$ and $qM$.
The definition of ${b\over M}$ comes from equating the quantum mechanical
angular momentum $\hbar \sqrt{l(l+1)}$ with the classical angular momentum
$\mu v_{\infty } b $. Finally, note that the slow motion approximation
$v_{\infty } \ll 1 $ is equivalent to
$$\hbar q \ll \mu \eqn\slow $$
that is, the (quantum mechanical) particle is nonrelativistic.
Although we don't have sufficient control of the solutions for the
eigenfunctions, to actually match the two asymptotic regions, we can
determine the scaling with $qM$ of the coefficients $C_{ql}$ and $S_{ql}$
in the low energy limit. Then in the next section we will use different
techniques to get approximate forms for $C$ and $S$ in essentially
all $qM$ and $l$ regimes. Now, suppose $qM\ll 1$ . Then for
$|R|<1/q$, the solutions to \radial\ are (approximately) independent of
$q$. The solution in the region $M<R<1/q$ inherits an overall scaling
dependence on $qM$ from the large $R$ solution \rbig . Hence this same scaling
with $qM$ is passed onto the solution in the region $-1/q < R< -M$,
which can then be matched onto the large negative $R$ region.
One finds, for $qM\ll 1$ and all $l$,
$$\eqalign{ S_{ql} \simeq & \ (-i)^{l+1} + \  O(qM)^{2l+2}  \cr
C_{ql} \simeq & \  {{(-i)^{l+1} \pi \gamma _l }\over {2^{\nu +l}
\Gamma (\nu + {1\over 2}) \Gamma (l+{3\over 2})}} (qM)^{\nu +l+1}
\cr}\eqn\coeff $$
where $\gamma _l $ is an unknown coefficient which is independent of
$q$. For large $l$ then, the capture coefficient is
$$|C_{ql}|^2 \simeq \  {{\pi ^2 e^{6l}}\over {l^2 2^{10l-1}}}
|\gamma _l |^2
 \left( {qM\over l}\right) ^{6l+3} \  , l\gg 1 \  ,qM\rightarrow 0
\eqn\capture $$
and
$$|C_{q0}|^2 \propto (qM)^2 \  \  ,for \  l=0. $$
(A similiar analysis in the case of a Klein-Gordon field scattering off a
Schwarzchild background yields $|C_{ql}|^2 \propto (qM)^{2l+2}$.)
Therefore at low energies the only significant capture occurs for
$l=0$ waves.

\chapter{Approximate Methods for finding $C_{ql}$}

We now turn to the calculation
of the capture coefficient. When the energy of the particle is well
over the potential barrier one can use the Born approximation, and the
WKB
approximation when the particle is well under the barrier,
for $l\gg 1$. For $l=0$ and low energies, one can use another approximation
where the potential is replaced by a delta-function.
For a given $
l$ and $qM$ one has to solve a cubic equation to decide if the
particle is over or under the barrier.
In the case $\mu =0$ this simplifies.
If $l=0$, then the particle is over the barrier if $qM> .23 $.
For $l\geq 1$, the particle is over the barrier --
and hence primarily captured-- if $qM> {2\over 5}
\sqrt{l(l+1)}$, which is equivalent to ${b\over M}< 2.5$.
(Recall that in the
classical case, the coalescence occurs if the impact parameter is less
than $2.5M$.)

\centerline{a) Low Energies}

In the small $qM$ regime one looks for wave functions of the form $|S'|^{
-1/2} e^{iS}$. In the first WKB or adiabatic approximation, $S=-\int ^R
(q^2 -\  V)^{1/2} dR' $, for a wave propagating in from $R=+\infty $. This
neglects terms of order $|{{\partial V}\over {\partial R}}||q^2 -V|^{-3/2}
$, which are small except near the classical turning points
$R_a$ and $R_b $ where $q^2 =V$. Further, for the WKB approximation to be
valid, the width of the potential $R_b -R_a $ must be large compared to the
wavelength of the incident particle. For $l\not= 0$ and small incident
energies, $qM\ll 2/5 \sqrt{l(l+1)}$, then $R_a \simeq -2l(l+1)/R^2 $ and
$R_b \simeq l(l+1)/R^2 $, so the WKB approximation is valid if $\sqrt{l(l+1
)} \gg 2 $. For $l=0$ WKB is never valid; one can't fit several wavelengths
in under the barrier.
Then
using the standard WKB matching formulae one finds the
capture coefficent for the case when the particle is well under the
barrier, and $\sqrt{l(l+1)}\gg 2$,
$$|C_{ql}|^2 = \ 4\left( 2\theta + {1\over {2\theta }}\right)^{-2}
\eqn\captwo $$
where
$$ln \theta =\  \int_{R_a}^{R_b} dR
(V-q^2 )^{{1\over 2}} .$$
In the intervals $R_a <R<-M$ and $M<R<R_b $, $V$ goes like $1/R^2$,
and the integration can be done exactly. In the remaining region
of integration,
$q^2$ can be neglected compared to $V$, and the value is approximately
$\int _{-M}^M dR \sqrt{V} \simeq l $ (which actually turns out to be a fair
estimate from doing the integral numerically). Using the values for
$R_a $ and $R_b$ given above, and keeping terms of leading order in
$qM/l$, we find
$$|C_{ql}|^2 \approx {1\over {\theta ^2 }}\approx ({{qm}\over l})^{6l+3}
{{e^{4l}}\over {2^{10l+5}}},\ \  l\gg 1 ,\  {{qM}\over l}\ll 1 .
\eqn\capthree $$

	The dependence on $qM$ in
\capthree\ agrees with the expression which was derived from
scaling arguments \impact (Indeed,
one can now approximate the unknown coefficient
$|\kappa _l |^2  \approx l^2 e^{-2l}/\pi ^2 2^6 $.)  The
important point is that the capture coefficient is very small; performing
the integral numerically one
finds the capture coefficient equals zero to 5 significant figures, in
ranges where the WKB approximation is reliable.

	For $l = 0$ and energies
below the potential barrier we must use a different technique.  In this
case the wavelength of the
incident wave is much greater than the width of the potential, as long as
$qM \ll 1$, that is, M is
small compared to the
deBroglie wavelength $\sqrt{2\mu E} /\hbar$.  (In this regime there
is no particular additional
constraint for consistency
with the slow velocity requirement \slow .  It is
sufficient, that $\mu M/\hbar  = \mu M/m_{Planck}^2 $
is less than or of order one.)  This long wavelength wave is not
sensitive to the details of the
potential and the potential in \potone\ can be replaced by a delta-function,
$V\rightarrow  LV_0\delta (R)$, with strength fixed by
$LV_0 = \int dR V(R)$.

	This problem can be solved, and the capture coefficient is
$$|C_{q0}|^2 =(1+({{LV_0 }\over{2q}})^2 )^{-1} \approx
(1+({{10^{-2}}\over{qM}})^2 )^{-1} ,\  l=0,\  qm\ll 1 \eqn\capfour$$
It is simple to estimate $LV_0$
to be a few times $10^{-2}/M^2$, and numerical integration gives
$LV_0  = 2\cdot 10^{-2} /M^2$.
As $qM\rightarrow 0$ the capture coefficient goes like
$(qM)^2$, which agrees with the previous result derived
from scaling arguments.

	Classically, if
the two black holes approach each other with zero angular momentum, they
always coalesce.  In the
quantum mechanical system, for incident energies of $qM = .1, .01$ and
$.001$, the $l = 0$ mode has a
capture coefficient of .99, .50 and .001 respectively.  The behavior
interpolates between almost
complete capture, as one expects for a particle, to complete scattering
to the asymptotic free
region, as is characteristic of a wave.  At very long wavelengths the wave
barely "sees" the black hole.

	In summary, for
incident energies under the barrier we can easily compute the captured flux
from \capfour , and the
higher angular momentum waves are almost completely scattered.  Only the
$l = 0$ mode has a non-negliable
capture coefficient.  This is not too surprising since an angular
momentum barrier is hard to
tunnel through.  (\capfour\ can also be used for incident energies over
the barrier, $.23 < qM < 1$,
where it gives a capture coefficient of essentially unity.  This agrees
with the results of the high energy approach discussed next.)

\centerline{b) High Energies}

	For energies high
compared to the height of the potential the simplest approximation is the
Born approximation, which appears
to be sufficient here.  To this end, we rewrite the
eigenfunction equation as an integral equation

$$\xi (R) = \xi ^0 (R) + \int dR' G(R, R') V(R')\xi (R') , \eqn\green $$
where $\xi ^0 (R)$ is any
solution to the equation with the potential set to zero, and G(R, R') is the
Green function for that equation.
For the no-flux-out-of-the-horizon boundary conditions as
before one takes $\xi ^0 = e^{-iqR}$ and
$$G(R, R') ={{ e^{iq(R-R')}}\over {2iq}} ,\  for R > R'$$
$$  G(R, R') = {{e^{-iq(R-R')}}\over {2iq}} ,\  for R < R' .$$
Substituting into \green , we
have $\xi$ as the sum of an ingoing and an outgoing wave, and hence can
deduce the scattering and
capture coefficients, S and C.  It is simple to check that this
approximation does not
conserve probability.  However, if one computes S and C to second order,
probability is conserved.
Furthermore, only C gets a correction at this order.  S is already correct
to second order, so it
suffices to compute S to first order and derive $| C |^2$ from
$| C |^2 = 1 - | S |^2$.

	The range of validity of the Born approximation is
$$qM \gg .23 , \ if \  l = 0$$
$$qM \gg (2/5)\sqrt{l(l+1)} \ if\  l > 0 .$$
In these ranges one has
$$|S_{ql}|^2 ={1\over {4q^2 }}|\int dR V(R) e^{-2iqR} |^2 \eqn\born $$
The integral is approximately zero
where the Born approximation is valid:  the width of the
potential at half-maximum is
about 3M, which means that for $qM \gg 1$, the exponential in the
integrand is rapidly oscillating
and different contributions cancel.  We computed the integral
numerically and indeed one
finds that, for $qM\geq 10$, once the particle is over the barrier, the
scattering coefficient is zero to five significant figures.

\centerline{c) Summary of Scattering Behavior}

	For $qM \geq 10$ we see
particle like behavior.  Recall that the condition for the wave to be over
the barrier, and hence captured, is
$${b\over M}={{\sqrt{l(l+1)}}\over {qM}} \leq 2.5 ,$$
which is the same as the
classical condition for coalescence.  For high enough energies the
discreteness of l is no
longer important; a sketch of the capture coefficient vs. l looks very much
like the plot of capture vs. a
continuous b/M.  Thus, for high energies, $| C |^2 \approx 1$ for $b/M < 2.5$.
For $b/M \geq 3$, where we know
the WKB approximation is valid, $| C |^2 \approx 0$.  Indeed, $| C |^2$ is
probably close to zero
at an impact parameter closer to 2.5 than 3.

	At intermediate energies,
$.23 \leq qM \leq 10 ,\  q^2$ is still above the
potential barrier for some $l-$values.  As $l$, or $b$,
is increased, $| C |^2$ decreases from (nearly)
unity to (nearly) zero.  The transition from the Born regime, where
$| C |^2 \approx 1$, to the WKB, where $| C |^2 \approx 0$,
is not as sharp as in the high energy case.  For
$qM = .5$ this transition occurs over a change in b of about 14M, compared
to M/2 above.
	For $qM < .23$, we see
transition to wave-like behavior, as previously discussed.  The wave
is almost completely
scattered, the captured flux going to zero like $(qM)^{6l+3}$
for large l and as $(qM)^2$ for $l = 0$.

	Apriori, one might expect
more back scattering in the over-the-barrier case and more
capture in the under-the-barrier
case than we have found; except for the lowest angular
momentum modes, the
transition from a capture coefficient of 1 to 0 is
quite abrupt in b/M.  This is due to the "featurelessness" of
the black hole potential.  There is only one length scale, M, which
determines both the height
and the width of the potential.  Typically in scattering problems there are
two independent parameters to vary.

	One can also think
of the over-the-barrier condition as fixing q and letting M increase, since
the analysis is valid in a
strong gravity regime.  Then, as one expects for a black hole, once particle
flux is in, it never gets
out.  Note that because of the slow motion approximation \slow , in the
regime $qM > 1$ one can only consider black holes with masses such that
$${{\mu M}\over {\hbar}}+{{\mu M}\over {m_{planck}^2}} >1, \eqn\uncer $$
in the case when the particle is over the barrier.

\chapter{Discussion}

At this point the
most interesting question is, what has been quantized?  When solving the
classical problem one
first solves for the trajectories of the black holes which immediately gives
the diagonal metric components
and the time component of the Maxwell potential.  Then one finds
(at least in principle)
the other field components by solving given constraint equations, in which the
black hole positions and
velocities appear as source terms.[2]  The complete solution can be found in
this sequence of steps because of the slow motion approximation.

	In the present
calculation we have quantized the distance between the two black holes.
Instead of having a solution
$\bf{r}(t)$ which is this distance as a function of time, one now has an
amplitude $\Psi (r,t)$.

This is fundamentally different
from calculations in which "gravitons" are quantized as
linear perturbations off some
fixed classical background (typically Minkowski or Robertson
Walker).  In the charged
black hole problem, one of the degrees of freedom of the full metric has
been quantized, not just a
fluctuation.  To see this, just write the spacetime metric \metrictwo in
coordinates such that one
of the black holes is at the origin and the other is at comoving coordinate
z, so that one of the
metric components is $r(t) = a(t)z$, the distance between the two black holes.

	On the other hand,
we have not allowed for quantum fluctuations in "directions in the space
of metrics" other than this
single function.  In a fully quantum mechanical theory one must allow
all the field components to vary
from the classical values.  The situation is similar to the
quantization of a nonlinear sigma
model.  Suppose we have a classical theory in D dimensions (or
with D fields) with a strong
potential that approximately constrains the dynamics to a D - 1
dimensional submanifold with
metric $g_{ab}$.  This leads us to model the system by the sigma model
$L = g^{ab} \nabla _a \phi \nabla _b \phi$, where $\phi$ is
constrained to take values on the D - 1 dimensional submanifold.
Quantizing the sigma model
also misses out on quantum fluctuations in directions off the
submanifold.  One expects
that quantizing the reduced system is a good approximation to the full
system, if the constraining potential
is sufficiently strong.  In the black hole problem this
corresponds to the slow
motion approximation remaining good.

	An alternative way to say
this, is that the approximation made here is the same as in "mini
superspace" models.  In these
quantum cosmology models, the wave function is taken to depend
on only one metric component, namely the scale factor.

	We also note that in
deriving our black hole Hamiltonian, $H_{eff}$, the classical equations of
motion have been inserted
into the action.  Now, it was checked by Eardly that to derive the
classical effective Hamiltonian,
one could either work exclusively with the equations of motion, or
with the action and
equations of motion for some of the fields, as described earlier.  However,
since all field configurations
contribute to a functional integral, one expects that use of the classical
equations for $N_i$ and $A_i$ also
leads to differences from a full quantum theory.  Again, this is similar
to the sigma model approximation discussed above.

	Next, what about the other
fields in the problem?  We suggest that the prescription for
recovering the metric and
electro-magnetic fields from the wave function for the black hole
coordinates is similar to
the prescription for recovering the states of Schroedinger's cat in that
famous demonstration of quantum
mechanics.[11]  That is, observed macroscopic states occur with
probabilities predicted by quantum
mechanics, but are not superpositions of multiple quantum
mechanical states.  Suppose
for particular initial conditions of the wave function there is a
scattering coefficient,p.  Then
at late times, according to this prescription, one observes fields due
to two widely separated
black holes with masses $m_1$ and $m_2$ with probability p, and the set of
fields corresponding to one
large black hole with mass M, with probability 1 - p.  That is, one
predicts that certain field configurations occur with certain probabilities.

	An alternative prescription
would be to couple the classical fields to the expectation value of
the charge operator.  Then,
e.g., x would be the sum of two Coulombic pieces, with widely
separated poles, one with
strength $m_1 + m_2 (1 - p)$ and the other with strength $m_2 p$.  That is, the
resulting fields are prescribed
deterministically, and correspond to sources where "part of the black
holes coalesced and part
scattered", rather than different classical configurations occurring with
different probabilities.

	Clearly in the gravitational problem
one can only argue by analogy and according to what
we observe at accessible
energies, to choose the theory for specifying the classical fields.  At this
point in our understanding of
quantum gravity, we leave the choice to the sense and sensibility of
the reader.

\ack
We would like to thank Doug Eardley and Gary Horowitz
for frequent thoughtful and helpful discussions.

\ref{G. Gibbons and N. Manton, {\it Nucl. Phys. }B 274, 183 (1986).}
\ref{R.C. Ferrell and D.M. Eardley, {\it Phys. Rev. Lett. }59, 1617 (1987).
	R.C. Ferrell and D.M. Eardley,
"Slowly Moving, Maximally Charged Black Holes", In:
{\it Frontiers in Numerical
Relativity}, ed. C. Evans, L. Finn and D. Hobill, (Cambridge
University Press, 1989), p. 27ff.}
\ref{ R.C. Ferrell and D. Eardley, "Slow
Motion Interactions of Maximally Charged Black Holes",
NSF-ITP Preprint 88-000.}
\ref{ S.D. Majumdar, {\it Phys. Rev. }72, 39 (1947).}
\ref{ A. Papapetrou, {\it Proc. Irish Acad. Sci., Sec. A }51, 191 (1947).}
\ref{N.S. Manton, {\it Phys. Lett.} B 110,
54 (1982); G. Gribbons and P. Ruback, {\it Phys. Rev. Lett.}
57, 1492 (1986).}
\ref{ We are indebted to
Gary Horowitz for providing us with some notes of R. Geroch and R.
Wald which helped clarify this point.}
\ref{ L. Parker, {\it Phys. Rev. }D 19, 438 (1979).}
\ref{ B. Harrison, K. Thorne, M.
Wakano, J. Wheeler, {\it Gravitation Theory and Gravitational
Collapse} (Univ. of Chicago Press, Chicago, 1965).}
\ref{ K. Gottfried, {\it Quantum Mechanics}, Vol.
I, (Benjamin/Cummings, London, 1974).}
\ref{ See, e.g., Schroedinger's
article in {\it Quantum Theory and Measurement}, ed. J.A. Wheeler and
W.H. Zurek (Princeton University Press, Princeton, 1983).}
\refout
\end